\documentclass[prl, aps, 12pt, showkeys, showpacs,tightenlines] {revtex4}
\usepackage{epsfig}
\usepackage{hyperref}
\begin{document}
\title{Entropy Production and Thermal Conductivity of A Dilute Gas}
\author{Yong-Jun Zhang}
\email{yong.j.zhang@gmail.com}
\affiliation{Science College, Liaoning Technical University, Fuxin, Liaoning 123000, China}

\begin{abstract}
It is known that the thermal conductivity of a dilute gas can be derived by using kinetic theory. We present here a new derivation by starting with two known entropy production principles: the steepest entropy ascent (SEA) principle and the maximum entropy production (MEP) principle. A remarkable feature of the new derivation is that it does not require the specification of the existence of the temperature gradient. The known result is reproduced in a similar form. 
\end{abstract}
\maketitle

\section{Introduction}
Thermal conductivity, $\kappa$, is a coefficient that appears in Fourier's law \cite{Fouriers_law, review1,review2,review3,wang},
\begin{equation}\label{jv}
        \mbox{\boldmath $j$}=-\kappa\mbox{\boldmath $\nabla$} T,
\end{equation}
where \mbox{\boldmath $j$} is the heat current density and $T$ is the temperature. For a dilute gas, the thermal conductivity can be derived by using kinetic theory \cite{gas,book2}, with the result
\begin{equation}\label{kd}
	\kappa=\frac{1}{3}ncv\lambda,
\end{equation}
where $n$ is the number density, $c$ is the molecular heat capacity, $v$ is the molecular average velocity and $\lambda$ is the mean free path. This derivation is simple, but it requires one to specify the existence of the temperature gradient, \mbox{\boldmath $\nabla$}$T$. We know that, along the direction of \mbox{\boldmath $\nabla$}$T$, the average molecular energy increases at rate $c$\mbox{\boldmath $\nabla$}$T$. Also, for a given direction, a molecule needs to travel an average distance $\lambda_z$  between two successive collisions. Thus one can simply consider along the \mbox{\boldmath $\nabla$}$T$ direction a layer of dilute gas with thickness $2\lambda_z$. Such a dilute gas can be viewed as two layers, each having height $\lambda_z$, and the two layers would exchange molecules. For each pair of molecules to be exchanged, the amount of energy that is exchanged is twice $\lambda_zc\mbox{\boldmath $\nabla$}T$. During unit time, across unit cross-sectional area, the number of molecules to be exchanged is $nv_z$. Therefore the heat flux density is $nv_z\lambda_zc\mbox{\boldmath $\nabla$}T$. By using the relations $\lambda_z=\frac{1}{\sqrt{3}}\lambda$ and $v_z=\frac{1}{\sqrt{3}}v$, the thermal conductivity Eq. (\ref{kd}) is obtained.

We shall present a new derivation. The new derivation does not require one to specify the temperature gradient. Instead, it uses the entropy production. Here, two aspects of the entropy production will be used: (1) for a given state, it evolves to a direction of the steepest entropy ascent; (2) among many candidates, the actual steady state has the maximum entropy production. In this paper, we call the first the steepest entropy ascent (SEA) principle and the second the maximum entropy production (MEP) principle. The name of "SEA" is originally introduced by Beretta \cite{Beretta3,Beretta32,Beretta4,Beretta5}. The MEP principle has been used by Paltridge \cite{Paltridge1, Paltridge2, Paltridge3} to study the Earth's climate.  There exist many studies of entropy production theory itself; see, for example \cite{Ziegler,MEPP,Dewar1,Dewar2,Dewar_book,Jaynes1,Jaynes2,Grinstein,Bruers,Niven,Croatica,last}. This paper provides an application.  

In order to derive the thermal conductivity of dilute gas, we shall consider a dilute gas that is fixed in between two thermal baths at different fixed temperatures. The dilute gas would carry a steady heat flux from whose expression the thermal conductivity can be extracted. We shall determine the form of the steady heat flux by using entropy production. In order to do that, first of all, we shall obtain the entropy of the dilute gas.

\section{Entropy of Dilute Gas}
In this section, we derive the entropy of a dilute gas. Let the dilute gas consist of $N$ molecules, and let $N$ be large.
The entropy depends on the heat flux, which in turn is related to the microscopic configurations of each molecule, including in which direction it transfers energy, how much energy it transfers, and how often it transfers energy.

Concerning the direction, we consider that each molecule transfers energy either upward or downward. Thus the probability that $k$ molecules transfer energy upward obeys the binomial distribution $B(N,\frac{1}{2})$, which can be approximated by the normal distribution $\mathcal{N}(\frac{N}{2},\frac{N}{4})$:
\begin{equation}\label{Pk}
        P(k)=\frac{1}{\sqrt{\pi \frac{N}2}}\exp\left({-\frac{(k-\frac{N}{2})^2}{\frac{N}{2}}}\right).
\end{equation}
  
The height of the dilute gas is denoted by $\Delta z$. We temporarily set $\Delta z$ equal to $2\lambda_z$ so that the dilute gas can be viewed as two layers, each of which has height $\lambda_z$ and consists of $N/2$ molecules. For the bottom layer, if $k_1$ molecules transfer energy upward, the number of molecules that transfer energy downward would be $N/2-k_1$, while, for the top layer, $k_2$ molecules transfer energy upward and $N/2-k_2$ molecules transfer energy downward. Here, $k_1+k_2$ actually is $k$. Thus the energy transferred from the bottom layer to the top layer is proportional to   
\begin{equation}
        k_1-(\frac{N}{2}-k_2)=k-\frac{N}{2}.
\end{equation}

Concerning how often a molecule transfers energy, on average a molecule transfers energy once every time interval ${\lambda}/{v}$, since on average a molecule collides once during that time interval. Concerning how much energy a molecule transfers, we denote the average of that by $\varepsilon$. We know that, when two molecules collide, the energy transferred is a little bit less than the energy difference between them. The average of that energy difference is comparable with the molecular energy distribution width, which is further comparable with the average molecular energy $cT$. So $\varepsilon$ is comparable with $cT$, but a little bit less.

Thus the heat flux is 
\begin{equation}\label{discussion_epsilon}
	J=\frac{\varepsilon v}{\lambda}(k-\frac{N}{2}).
\end{equation}
By using Eq. (\ref{Pk}), we see that 
\begin{equation}
	P(J)=\frac{1}{\sqrt{\pi \frac{N}{2} }}\exp\left(-\frac{2\lambda^2}{N \varepsilon^2v^2}J^2\right).
\end{equation}
This equation reflects the fact that, even for a dilute gas that is isolated, a fluctuating heat flux $J$ may still occur with certain probability $P(J)$.

We know that
\begin{equation}
        \Omega(J)\propto P(J),
\end{equation}
where $\Omega(J)$ is the number of the microscopic states that are associated with $J$. Using the Boltzmann formula
\begin{equation}
	S(J)=k_B\ln\Omega(J),
\end{equation}
we write for the entropy of the dilute gas 
\begin{equation}\label{dilute_gas_entropy}
	S(J)=S_0-\frac{2k_B\lambda^2}{N \varepsilon^2v^2}J^2=S_0-\frac{1}{2}\frac{\Delta z}{A}\frac{3k_B}{n \varepsilon^2v^2}J^2,
\end{equation}
where $S_0$ is introduced to denote the entropy of the dilute gas when it is in the equilibrium state, $N$ is equal to $2nA\lambda_z$, and $A$ is the cross-sectional area. Here, $\Delta z$ has been temporarily set equal to $2\lambda_z$. But notice that $\Delta z$ can be extended to any value, because entropy is an additive physical quantity. Thus we write the general form of the entropy of a dilute gas with any height $\Delta z$ as 
\begin{equation}\label{system_entropy}
	S(J)=S_{0}-\frac{1}{2}C_2J^2,
\end{equation}
with
\begin{equation}\label{C_2}
	C_2=\frac{\Delta z}{A}\frac{3k_B}{n \varepsilon^2v^2}.
\end{equation}

We may also obtain the entropy per unit volume, $s=\frac{S}{A\Delta z}$, with respect to heat flux density $j=J/A$,
\begin{equation}\label{sj} 
	s(j)=s_{0}-\frac{1}{2}\frac{3k_B}{n \varepsilon^2v^2}j^2.
\end{equation}
A similar formula is also proposed by Lebon {\it et al.}  \cite{math, math2}, in the form 
\begin{equation}\label{sTq} 
	\rho s(T,\mbox{\boldmath $q$})=\rho s_0(T)-\frac{1}{2}\tau(\lambda T^2)^{-1}\mbox{\boldmath $q\cdot q$},
\end{equation}
where $\rho$ is the mass density, $s$ is the entropy per unit mass, \mbox{\boldmath $q$} is the heat flux density vector, and $\tau$ is the relaxation time; $\lambda$ has a different meaning here: it means the thermal conductivity. In the end, we will obtain the thermal conductivity (\ref{kappa2}), by using which one can see that Eqs. (\ref{sj}) and (\ref{sTq}) are the same.

\section{Entropy Production and Thermal Conductivity of Dilute Gas}
When the dilute gas is fixed between the two baths at different fixed temperatures, a steady heat flux will flow through it. In order to determine the steady heat flux, it seems that, for every possible heat flux, we need to examine all possibilities of how it evolves. Fortunately, it will turn out that, if we use the SEA principle in conjunction with the MEP principle, for a given heat flux we only need to examine two specific possibilities.

One possibility is that the given heat flux $J$ is steady. For this possibility the entropy production would be
\begin{equation}\label{steady_EP}
        \sigma=(\frac{1}{T_1}-\frac{1}{T_2})J,
\end{equation}
where $T_1$ and $T_2$ are the fixed temperatures of the baths and $T_1<T_2$. Here, the heat flux flows from the bath at the higher temperature $T_2$ to the bath at the lower temperature $T_1$.

The other possibility is that the given heat flux $J$ starts to relax at the maximum rate,
\begin{equation}\label{relax}        
	J_{R}(t)=J\exp\left(-\frac{t}{\tau}\right),
\end{equation}
where $\tau$ is the minimum possible relaxation time, which is approximately equal to $\lambda/v$.
To see why the relaxation is exponential and $\tau\approx \lambda/v$, let us study an example. In the example, let the height of the dilute gas be $3\lambda_z$, so that it can be viewed as three layers, each having height $\lambda_z$. The two end layers are in  contact with the thermal baths and their temperatures are therefore different. Through all three layers a heat flux flows steadily. Now, suppose that the thermal baths are suddenly detached. The heat flux of the two end layers will cease immediately while the heat flux of the middle layer will relax gradually. Thus the two end layers act as new thermal baths but their temperatures are brought closer and closer by the heat flux flowing through the middle layer. The closer the two temperatures are, the smaller the heat flux becomes. The smaller the heat flux becomes, the slower the rate that the two temperatures get close will be. Thus both the temperature difference and the heat flux decrease exponentially. Concerning $\tau$, one needs to consider the microscopic mechanism of the heat flux.
  One may view the middle layer as a collection of individual molecules bouncing between the two end layers. On average, each bounce takes time $\lambda/v$ and transfers energy comparable with the molecular energy difference between the two end layers. Thus just a few bounces can bring the two temperatures close. So the relaxation time $\tau$ is comparable with $\lambda/v$ but a little bit larger.
A dilute gas having height larger than $3\lambda_z$ can be viewed as many layers, each having height $\lambda_z$. Then the discussion would be the same, but each layer except for the two end layers plays two roles: it carries a relaxing heat flux and it acts as a thermal bath. Thus all middle layers are correlated, and such correlation makes the relaxation time larger. But we are only concerned with the instant at which the heat flux starts to relax. At that instant, the correlation has not yet taken effect, and thus, in the limit $t\to 0$, $\tau$ is the same for a dilute gas having any height.

Using Eq. (\ref{system_entropy}), we see that
\begin{equation}
	S(t)=S_{0}-\frac{1}{2}C_2J_{R}(t)^2=S_{0}-\frac{1}{2}C_2J^2\exp\left(-2\frac{t}{\tau}\right),
\end{equation}
and
\begin{equation}
	\sigma(t)=\frac{dS(t)}{dt}=\frac{C_2}{\tau}J^2\exp\left(-2\frac{t}{\tau}\right),
\end{equation}
which, in the limit $t\to 0$, leads to
\begin{equation}\label{sigma_S}
	\sigma=\frac{C_2}{\tau}J^2.
\end{equation}
The reason why we use the limit $t\to 0$ is that we only need to know whether or not the given heat flux would start to relax. The relaxation process itself is irrelevant to our problem. (Also, in the limit $t\to 0$, $J_R(t)$ should be uniform in space and time.)

There is one issue to be addressed: the entropy production within each bath can be neglected if the baths are chosen to be perfect. By 'perfect' we mean that their thermal conductivity $\kappa$ is very large and their $C_2$ is very small. When their $\kappa$ is very large, the part of their entropy production arising in the form of Eq.(\ref{steady_EP}) can be neglected, because the temperature distribution within each bath would be uniform. When their $C_2$ is very small, the part of their entropy production arising in the form of Eq. (\ref{sigma_S}) can be neglected too. So we do not need to consider the entropy production of the baths. 

\begin{figure}[htbp]
  \begin{center}
    \mbox{\epsfxsize=12.0cm\epsfysize=10.0cm\epsffile{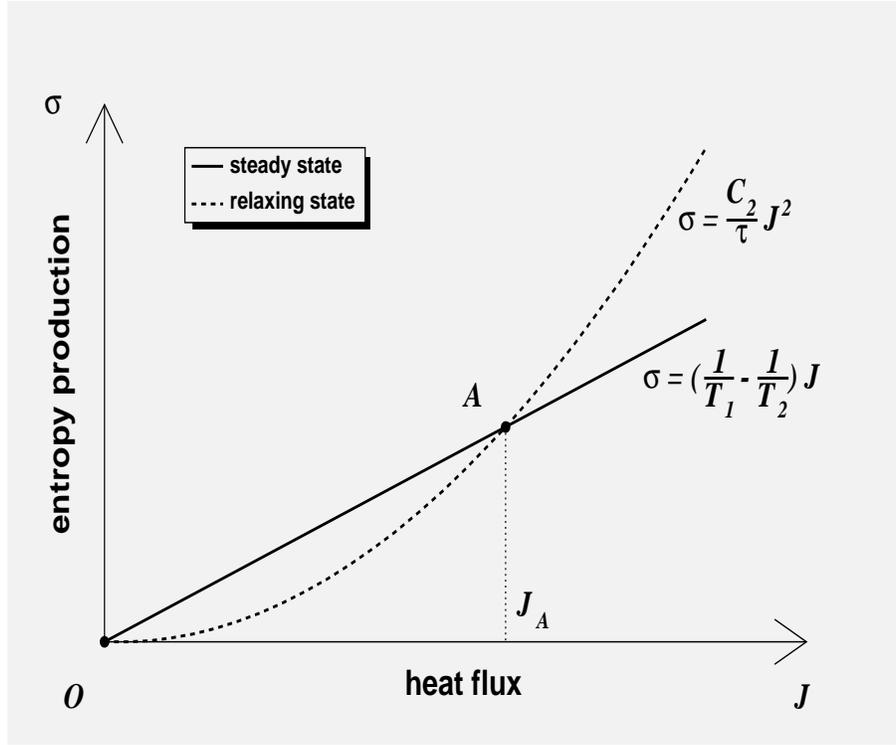}}
  \end{center}
\caption{Two entropy productions with respect to small heat flux. The solid line is the entropy production when the heat flux is steady. The dashed line is the entropy production when the heat flux starts to relax at the maximum rate. $J_A$ is the determined steady heat flux. For a heat flux $J>J_A$, it cannot be steady because in this range the solid line is not the highest, at least is lower than the dotted line. The steepest entropy ascent (SEA) principle indicates that a state can occur only if it is associated with the maximum entropy production. Thus the steady heat flux must be in the range $0\le J\le J_A$. Subsequently, according to the maximum entropy production (MEP) principle, we know that the steady heat flux must be $J_A$ because among all candidates it is associated with the maximum entropy production.\label{fig1}}
\end{figure}

The entropy productions of the two possibilities are plotted in Fig. \ref{fig1}. They are equal when $J=J_A$. We shall use first the SEA principle and then the MEP principle to determine that the steady heat flux must be $J_A$. According to the SEA principle, for a heat flux $J>J_A$, the heat flux cannot be steady because the entropy production would not be the maximum. (But it would not start to relax at the maximum rate either. Actually, the heat flux will start to relax at a smaller rate until $J=J_A$, and during the entire relaxation process the dilute gas will continue to exchange heat with the baths.) Thus the steady heat flux must be in the range $0\le J\le J_A$. Then according to the MEP principle, we see that the steady heat flux must be $J_A$ because, among all candidates, $J_A$ is the one that is associated with the maximum entropy production.

The steady heat flux in the figure is $J_A$, and therefore the steady heat flux is
\begin{equation}\label{J}
	J=\frac{\tau}{C_2}(\frac{1}{T_1}-\frac{1}{T_2}).
\end{equation}
Plugging Eq. (\ref{C_2}) in, we write for the heat current density 
\begin{equation}\label{j}
        j=\frac{J}{A}=\frac{\tau n \varepsilon^2v^2}{3k_B}\frac{1}{\Delta z}(\frac{1}{T_1}-\frac{1}{T_2}),
\end{equation}
which becomes, in the limits $\Delta z \to 0$ and $T_2\to T_1$,
\begin{equation}\label{jz}
        j=\frac{\tau n \varepsilon^2v^2}{3k_B T^2}\frac{d T}{d z}.
\end{equation}
From Eq. (\ref{jz}) and Eq. (\ref{steady_EP}), we see that the heat flux indeed flows from the bath at the high temperature to the bath at the low temperature. This is the same as the minus sign in the equation (\ref{jv}) indicated. Finally, the thermal conductivity of the dilute gas is extracted as
\begin{equation}\label{kappa2}
         \kappa=\frac{\tau n \varepsilon^2v^2}{3k_B T^2}.
\end{equation}
If one uses the approximations
\begin{equation}
        \tau\sim \frac{\lambda}{v}, \ \varepsilon\sim cT,\ k_B\sim c,
\end{equation}
one will find that the new result (\ref{kappa2}) is the same as the kinetic theory result (\ref{kd}). In fact, $\varepsilon$ is a little bit less than $cT$, as discussed before Eq. (\ref{discussion_epsilon}), 
while $\tau$ is a little bit larger than ${\lambda}/{v}$, as discussed after Eq. (\ref{relax}). Also, $c$ is a little bit larger than $k_B$; for example, for an ideal gas $c=\frac{3}{2}k_B$. So the two results may be different, but they should remain comparable with each other.

\section{Conclusion}


We have derived the thermal conductivity of a dilute gas by using a new approach of entropy production. The derivation is based on the fact that there exists an entropy production competition between at least two specific possibilities. One possibility is that the heat flux is steady. The other possibility is that the heat flux relaxes exponentially at the maximum rate. Then, by applying the steepest entropy ascent (SEA) principle, we have obtained the upper limit of the steady heat flux. Subsequently, by applying the maximum entropy production (MEP) principle, we have identified the actual steady heat flux. The resulting expression for the thermal conductivity is comparable with the result of kinetic theory. As an intermediate step, we have found that the non-equilibrium entropy of a dilute gas is lower than the equilibrium value by an amount proportional to the square of the heat flux.

{\bf Acknowledgment:} Work supported by Liaoning Education Office Scientific Research Project(2008288), and by SRF for ROCS, SEM.


\begin{thebibliography}{99}
\bibitem{Fouriers_law} J. Fourier, {\it The analytical theory of heat}, Dover Publ., New York, 1955, transl. by Alexander Freeman. Original: {\it The$\acute{\rm o}$rie analytique de la chaleur} (publ. 1822) p. 52.
\bibitem{review1} M. Michel, J. Gemmer and G. Mahler, {\it Int. J. Mod. Phys}. {\bf{ B 20}} (2006), 4855. \href{http://arxiv.org/abs/cond-mat/0611612}{arXiv:cond-mat/0611612} 
\bibitem{review2} M. Buchanan, {\it Nature Physics} {\bf 1} (2005), 71. 
\bibitem{review3} F. Bonetto, J. Lebowitz, and L. Rey-Bellet, {\it Fourier’s law: a challenge to theorists, Mathematical Physics 2000} (London: Imperial College Press), pp. 128-50. \href{http://arXiv.org/abs/math-ph/0002052}{arXiv:math-ph/0002052}
\bibitem{wang} Q. A. Wang, Astrophys. Space Sci. (2006) 305:273-281. 
\bibitem{gas} S. Brusch, {\it The kind of motion we call heat} (North Holland, 1976), p. 489. 
\bibitem{book2} Edited by Terry M. Tritt, {\it Thermal conductivity: theory, properties, and applications}, (Springer, 2004), p. 2.  
\bibitem{Beretta3} G.P. Beretta, Steepest Entropy Ascent in Quantum Thermodynamics, in {\it The Physics of Phase Space}, Edited by Y.S. Kim and W.W. Zachary, Lecture Notes in Physics, Vol. 278, Springer-Verlag, pp. 441-443 (1986).
\bibitem{Beretta32} G.P. Beretta, Phys. Rev. E {\bf 73}, 026113 (2006).
\bibitem{Beretta5} G.P. Beretta, International Journal of Quantum Information, Vol. 5, 249 (2007).
\bibitem{Beretta4} G.P. Beretta, Entropy {\bf 10}, 160-182 (2008). 

\bibitem{Paltridge1} G.W. Paltridge, Quart. J. Royal Meteorol. Soc. {\bf 101}, 475 (1975).
\bibitem{Paltridge2} G.W. Paltridge, Quart. J. Royal Meteorol. Soc. {\bf 104}, 927 (1978).
\bibitem{Paltridge3} G.W. Paltridge, Nature {\bf 279}, 630 (1979).
\bibitem{Ziegler} H. Ziegler, C. Wehrli J. Non-Equilib. Thermodyn. 12 (3) (1987) 229.
\bibitem{MEPP} L.M. Martyushev,V.D. Seleznev, Physics Reports {\bf 426}, 1 (2006).  
\bibitem{Dewar1} R.C. Dewar, J. Phys. A: Math. Gen. {\bf 36} (2003) 631-641.\href{http://arxiv.org/abs/cond-mat/0005382}{arXiv:cond-mat/0005382}
\bibitem{Dewar2} R.C. Dewar, J. Phys. A: Math. Gen. {\bf 38} (2005) L371.
\bibitem{Dewar_book} Edited by A. Kleidon, R.D. Lorenz, {\it Non-equilibrium Thermodynamics and the Production of Entropy}, (Springer, 2004). 
\bibitem{Jaynes1} E. T. Jaynes, Phys. Rev. {\bf 106} (1957) 620–630.
\bibitem{Jaynes2} E. T. Jaynes, Phys. Rev. {\bf 108} (1957) 171–190.
\bibitem{Bruers} Stijn Bruers, J. Phys. A: Math. Theor. {\bf 40} 7441 (2007). \href{http://arXiv.org/abs/0705.3226}{arXiv:0705.3226}
\bibitem{Grinstein} G. Grinstein, R. Linsker, J. Phys. A: Math. Theor. {\bf 40} 9717 (2007).
\bibitem{Niven} R.K. Niven, Phys. Rev. E {\bf 80}, 021113 (2009). \href{http://arxiv.org/abs/0902.1568}{arXiv:0902.1568}
\bibitem{Croatica} P. \u{Z}upanovi\'{c}, S. Botri\'{c}, D. Jureti\'{c}, Croatica Chemica Acta, {\bf 79} (3), 335-338 (2006).
\bibitem{last} G.P. Beretta, Reports on Mathematical Physics, Vol. 64, 139-168 (2009).
\bibitem{math} G Lebon, J Casas-Vazquez and D Jou,J. Phys. A: Math. Gen. {\bf 15} L565 (1982).
\bibitem{math2} J Casas-Vazquez and D Jou 1981 J. Phys. A: Math. Gen. {\bf 14} 1225.

\end{thebibliography}
\end{document}